\documentclass[conference]{IEEEtran}
\usepackage[english]{babel}
\usepackage[utf8x]{inputenc}
\usepackage[T1]{fontenc}
\usepackage{amsmath}
\usepackage{graphicx}
\usepackage{epsfig}
\usepackage[colorinlistoftodos]{todonotes}

\begin{document}
%

\title{On M2M Micropayments : A Case Study of Electric Autonomous Vehicles}

\author{\IEEEauthorblockN{Dragos Strugar\IEEEauthorrefmark{1}, Rasheed Hussain\IEEEauthorrefmark{1},
Manuel Mazzara\IEEEauthorrefmark{1}, Victor Rivera\IEEEauthorrefmark{1}, JooYoung Lee\IEEEauthorrefmark{1}, and Ruslan Mustafin \IEEEauthorrefmark{1}}
\IEEEauthorblockN{\IEEEauthorrefmark{1} Innopolis University, Russia, 
 Email: \{d.strugar,r.hussain,m.mazzara, v.rivera, j.lee, r.mustafin\}@innopolis.ru}
 }

\maketitle

\begin{abstract}
The proliferation of electric vehicles (EVs) has spurred the research interest in technologies associated with it, for instance batteries, and charging mechanisms. Moreover, the recent advancements in autonomous cars also encourage the enabling technologies to integrate and provide holistic applications. 
To this end, one key requirement for EVs 
is to have an efficient, secure, and scalable infrastructure and framework for charging, billing, and auditing. 
However, the current manual charging systems for EVs may not be applicable to the autonomous cars that demand new, automatic, secure, efficient, and scalable billing and auditing mechanism. Owing to the distributed systems such as blockchain technology, in this paper, we propose a new charging and billing mechanism for electric vehicles that charge their batteries in a charging-on-the-move fashion. To meet the requirements of billing in electric vehicles, we leverage distributed ledger technology (DLT), a distributed peer-to-peer technology for micro-transactions. Our proof-of-concept implementation of the billing framework demonstrates the feasibility of such system in electric vehicles. It is also worth noting that the solution can easily be extended to the electric autonomous cars (EAVs).   
\end{abstract}



%
\IEEEpeerreviewmaketitle

\section{Introduction}

Today's high-end vehicles are equipped with a plethora of sensors, communication, computation, and storage resources and thus-forth referred to as computers-on-wheels. Therefore, such vehicles are perfectly capable for communicating with each other and with the environment to implement Intelligent Transportation System (ITS) through Vehicular Ad hoc NETwork (VANET)\cite{wang2009vehicular}. 

However, amid the efforts for VANET realization, there has been pressing concerns about the side effects of the combustion engines that claim the majority of the transport today. To this end, Electric Propulsion Vehicle (EPVs or EVs) were introduced to address the environmental and economic issues. Over the years, the increase in oil prices also ignited the surge for alternative developments that are both cheaper and cleaner than the fuel-propelled engines. 

Technological solutions are essential to aid the regulations and to increase the efficacy of proposed solutions for billing and charging EVs. On the other hand, the recent advancements in deep learning, artificial intelligence, computer vision, and other related technologies turned a once science fiction phenomenon (autonomous or driverless car) into a reality. 
A fully autonomous car is a complex distributed system that combines various computation, communication, and storage domains with the focus on learning, artificial intelligence, vision and capability of decision-making. These cars have already traveled millions of miles as part of the test drive.   

The combination of environment-friendly transportation (EVs) and autonomous cars (Electric Autonomous Vehicle - EAV) is a perfect combination of desired intelligent, connected, safe and economical transportation system. 
The phenomenon of EAV can also be considered as a foundation for machine-to-machine economy where different transactions are needed, for instance, payments for ride-sharing, carpooling, renting, and charing-on-the-move, to name a few. Without loss of generality, this paper focuses only on the charging and billing aspects of EAV. To date, many efforts have been made to design charging and billing frameworks \cite{RUBINO2017438,Rezaeifar2017,Hussain2017}; however, the current solutions do not encompass all the requirements for M2M economy and EAVs. Furthermore, the deployment of these systems is still questionable.
To fill the gaps, we propose a practical and feasible framework architecture for charging and subsequently billing EAVs. We implement a proof-of-concept based on Distributed Ledger Technology (DLT) inspired by the widespread use of Blockchain technology. In our proposed architecture, we use IOTA-based payment system which is the implementation of DLT and offers numerous advantages such as scalability, proof of delivery, proof of payment, safety, efficiency, security, and privacy.

This paper is further organized as follows. We outline DLT and Tangle in Section \ref{background}, proposed framework in Section \ref{sec:architecture} and proof of concept in Section \ref{poc}. Section \ref{conclusions} concludes the paper. 

\section{Distributed Ledger Technology and Tangle}
\label{background}
Blockchain is a new decentralized architecture and DLT that has recently gained significant popularity. It represents an immutable ledger system implemented in a distributed way to eliminate the need for a trusted third-party in a transaction \cite{nakamoto2008bitcoin}. Blockchain increases the transparency of the whole system while maintaining the high level of its security. To date, considerable research has been conducted on the use-cases of DLT-related domains including Internet-of-Things (IoT). The main reasons for DLT being beneficial for IoT is the fact that IoT includes M2M value transactions, increased security and the automation of processes. However, technical limitations of the current blockchain architecture do not allow it to be broadly used in this industry, primarily due to high transaction fees and scalability issues \cite{medium2017introductiontoiota}. In most of the cases, machines need to make micropayments among themselves, rendering it infeasible to have transaction fees on micropayments. In addition, transaction confirmation times should be minimized for the network to be scalable and ready to accept new transactions. Recently, the Tangle has been introduced \cite{popov2016tangle} to address the problems of the traditional blockchain. Tangle is an underlying technology of the IOTA cryptocurrency\footnote{https://www.iota.org}. Based on Directed Acyclic Graph (DAG), Tangle serves as a distributed database with three additional advantages: it is scalable, involves no fees and allows offline transactions. These characteristics and its cryptocurrency make Tangle favorite to become the backbone of the emerging machine economy by allowing machines to trade resources (electricity in our case) and services with each other without the involvement of the intermediary.


\section{EAV Charging and Billing Architecture}
\label{sec:architecture}
In this section we describe the overall system architecture and its organization as well as the proposed technologies. 
The proposed framework consist of of three major layers, \textbf{physical}, \textbf{network}, and \textbf{services}. These layers are further elaborated below:


\subsection{Physical and Users Layer}
Physical layer is responsible for direct communication with the user, i.e. a vehicle, and it encapsulates all the hardware components embedded in the CS used for sensing and gathering information about the charging process. The core parts of this layer include energy calculator, main controller (MC), and Electric Vehicle Supply Equipment (EVSE). 

kWh meter is the main sensor in charging station and it keeps track of the amount of energy transferred to the consumer EAV. Depending on the price of the energy at any instant of time, the accumulated data is used to calculate the total price for the transferred energy. The benefit of such an approach is that consumers can get the amount of energy they have paid for. On the other hand, it is also important for the service provider to receive the amount of money equal to the amount of energy it has transferred to the consumer. 

Moreover, knowing the current energy rate can also help to adjust the maximum power consumed by the EAV, especially in the case of multiple CS using the same power source. The device used to perform such a task is the Smart EVSE \cite{smartevse2014cs}. Smart EVSE delivers the energy to EAV based on demand in a smart way to be both efficient and economical and thus-forth also contributes to the development of the Smart Grid \cite{SIANO2014461,medium2017flash}. Lastly, we use MC to transfer the information gathered by components of the physical layer to higher levels of the architecture through a special communication protocol designed for IoT, i.e., Message Queuing Telemetry Transport (MQTT). 

\subsection{Network and Communication Layer}
Since our scenario is composed of autonomous and dynamic decision-making CSs and vehicles, it makes more sense to establish a peer-to-peer (P2P) communication between these entities. Therefore, EAV and CS will have equal privileges to receive the energy and receive the payment, respectively \cite{yuan2016towards}. For P2P communication, we consider DLTs as a viable solution. More precisely, we choose IOTA's Tangle as the decentralized database that stores both data and value transactions. IOTA supports flash transactions where the value and the price execute simultaneously as opposed to the traditional blockchain technology. Furthermore, it also provides a wide range of possibilities for different networking technologies as well as support for both offline or online transaction of value and data, depending on the underlying protocol. The fact that our proposed architecture involves constant M2M communication with publish and subscription behavior, MQTT is the most suitable protocol for inter-device communication \cite{4554519}. In other words, MQTT ensures fast transfer of both data and value with high throughput.


In addition to the afore-mentioned features, the transactions between EAV and CS must be secure. To meet the requirements, Tangle uses a bi-directional payment channel, namely Flash Channel (FC), that conforms to the MQTT protocol by design \cite{medium2017flashch}. FC allows both consumers and service providers to get the service and get paid, respectively simultaneously. Moreover, FC also enables real-time streaming of transactions, and allows machines to parallelize the value and price, i.e. transmit data and value through any suitable protocol such as MQTT, and merge the results with the main IOTA Tangle. This phenomenon adds the flexibility to the architecture where the operational time is reduced by considerable amount because the communicating parties do not have to wait for transactions to be approved by the IOTA network. Accordingly, only the starting and ending transactions of the Flash channel will be published to the main network. 


Once the utility and its price have been exchanged between the communicating parties through FC, the final transaction denoted by $\mathit{t_f}$ is published on the Tangle. At this point, $\mathit{t_f}$ has to verify two non-validated transactions in the Tangle (selected randomly). It does so by performing computational work which is used to verify if the transaction is valid or not. This process proves the Proof-of-Work (PoW). Similarly, other transactions $\mathit{t_{k \neq f}}$, $\mathit{t_{l \neq k \neq f}}$ will find and verify $\mathit{t_f}$. In order for any transaction $\mathit{t_{x \neq f}}$ to be approved in the Tangle network, two incoming transactions $\mathit{t_y}$ and $\mathit{t_z}$ have to verify it using their PoW. This way, there is no need for miners, in fact, every user in the network is miner. In addition, unlike other blockchain technologies, the Tangle network benefits from new transactions as they help to verify two unapproved transactions which improve the scalability of the system \cite{popov2016tangle}. Hence through Tangle, FCs and MQTT protocol for message transfer, the network layer is able to transfer sensor data (electric energy in our case) from the physical layer to the services layer in a fast and secure way. 


\subsection{Services Layer}
The first two layers represent the basis for the Services Layer, i.e higher abstraction of the architecture. Here services to consumers and service providers are delivered. The nature of these services can be broad, for example:

\begin{itemize}
  \item Charging services for EAVs
  \item Data Insights for Service Providers
\end{itemize}


Charging services are envisioned as unmanned, i.e no humans involved in the process. AVs would, therefore, embed the application used for payment and would make use of Global Positioning System (GPS) and Artificial Intelligence (AI) to locate the most appropriate Charging Station. The application would also be responsible for initiating and taking care of the charging process, broadcasting the amount of electricity needed and paying for services provided using its IOTA wallet. Once the charging process is completed, the application would take care of closing the payment channel and record the new wallet balance on the main Tangle network. In this vision, EAV would have their own wallets, and all the process would be fully automated (M2M Economy).


Service providers benefit here from the possibility of accessing real-time data and make analyses. Considered the low-latency of NoSQL databases, the data collected by the Physical and Network Layers in the process of off-tangle Flash transactions could feed Machine Learning (ML) algorithms and generate graphical demos, predictions and live billing analysis.

\section{Proof of Concept}
\label{poc}
To prove the feasibility of our proposed scheme, we developed a proof of concept with traditionally available tools such as Raspberry Pi and a temperature sensor (Dallas Semiconductor DS18B20). We consider the case study of EAV charging its battery at a CS or on the move. 

\subsection{Case study: EAV charging and Billing}
The case study utilized in this paper is to build a proof of concept which is exhaustively described in~\cite{Strugar18}. When the vehicles' Battery Management System (BMS)\footnote{BMS is a combination of sensors, controller, computation and communication hardware with software algorithms designed to estimate the battery percentage of the EV \cite{5609223}.} indicates that the vehicle needs charging, the application embedded in the EAV starts searching for the most suitable CS nearby. 
First, EAV authenticates with the CS by fetching its wallet address and establishing a secure payment channel, i.e., FC. 
After successful authentication and establishment of FC, both parties have to deposit equal amount of IOTA into a multi-signature address controller by both parties. This step is mandatory as it would prevent parties from refusing to continue to sign the transactions. 
Networking layer processes the request and sends the data to the MC of the Physical Layer. When the CS gets this information, the charging process starts. Meanwhile, MC uses data from kWh meter to record the amount of electricity consumed and sent to the EAV's application via MQTT protocol. The consumer pays the calculated sum through the FC. The transactions are stored in a local (preferably NoSQL) database with low latency. This enables service providers to apply ML algorithms to this data and gain useful insights in real-time. 
When EAV receives the energy, FC takes the latest state and spreads the remaining tokens across transacting parties. After each party signs the final bundle, it is then attached to the main network, i.e., the Tangle. EAV's application records the end of the charging process, and sets its state to inactive. 



\subsection{POC Setup}
\begin{figure}
\includegraphics[scale=0.18]{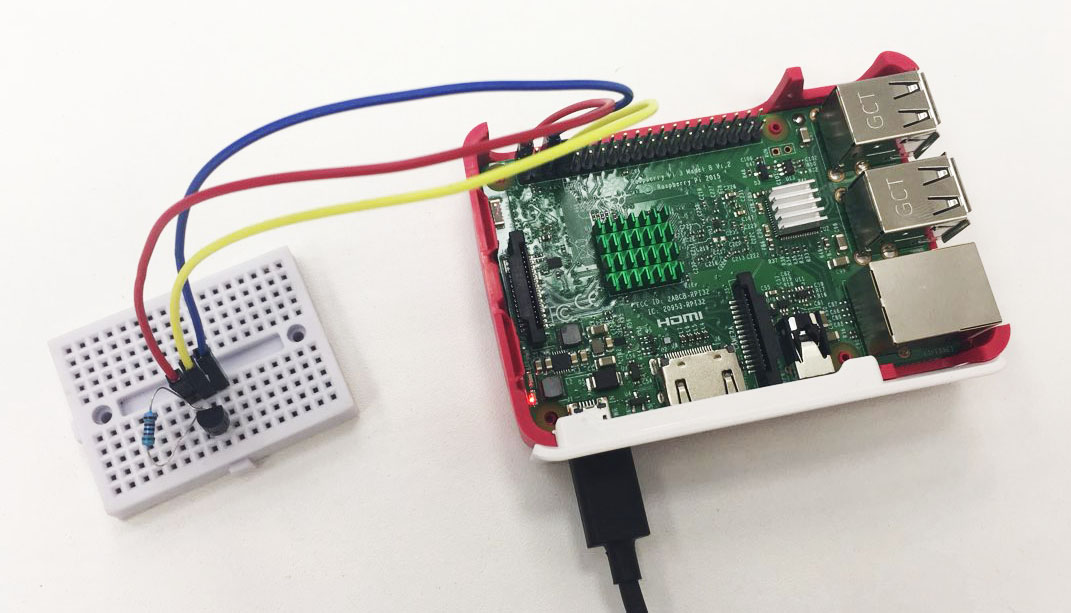}
\centering
\caption{Raspberry PI 3 with DS18B20 Temperature Sensor}
\label{fig:raspberrypi}
\end{figure}

To emulate the real-world charging and billing process, we chose to use the temperature sensor to emit real-time data. In future, the sensor could be replaced with module to measure the amount of electricity consumed (kWh meter) with minor adjustments. Figure \ref{fig:raspberrypi} shows our hardware setup. Raspberry PI 3 (PI), running Raspbian Stretch Lite operating system, is connected to the Dallas Semiconductor DS18B20 temperature sensor through General Purpose Input/Output (GPIO) pins. PI also contains the open-source and lightweight MQTT Broker, \textit{Mosquitto}, used to handle the publish/subscribe behavior. Temperature recordings are accumulated through our Python script \footnote{https://github.com/innoiota/python-server/blob/master/python\_server.py}
running inside the PI, which is also used for communication with other devices in our POC using MQTT protocol.

We also used Node.js\footnote{https://nodejs.org/} as the back-end server that serves three main purposes: listen to messages provided by the PI, make actual Flash transactions using IOTA Flash Library, and update the User Interface (UI) accordingly. In future, we plan to integrate a NoSQL database to have a better storage mechanism, which would allow the charging service providers to expand the set of services they provide (live billing analysis, reports, etc). 

\begin{figure}
\includegraphics[scale=0.35]{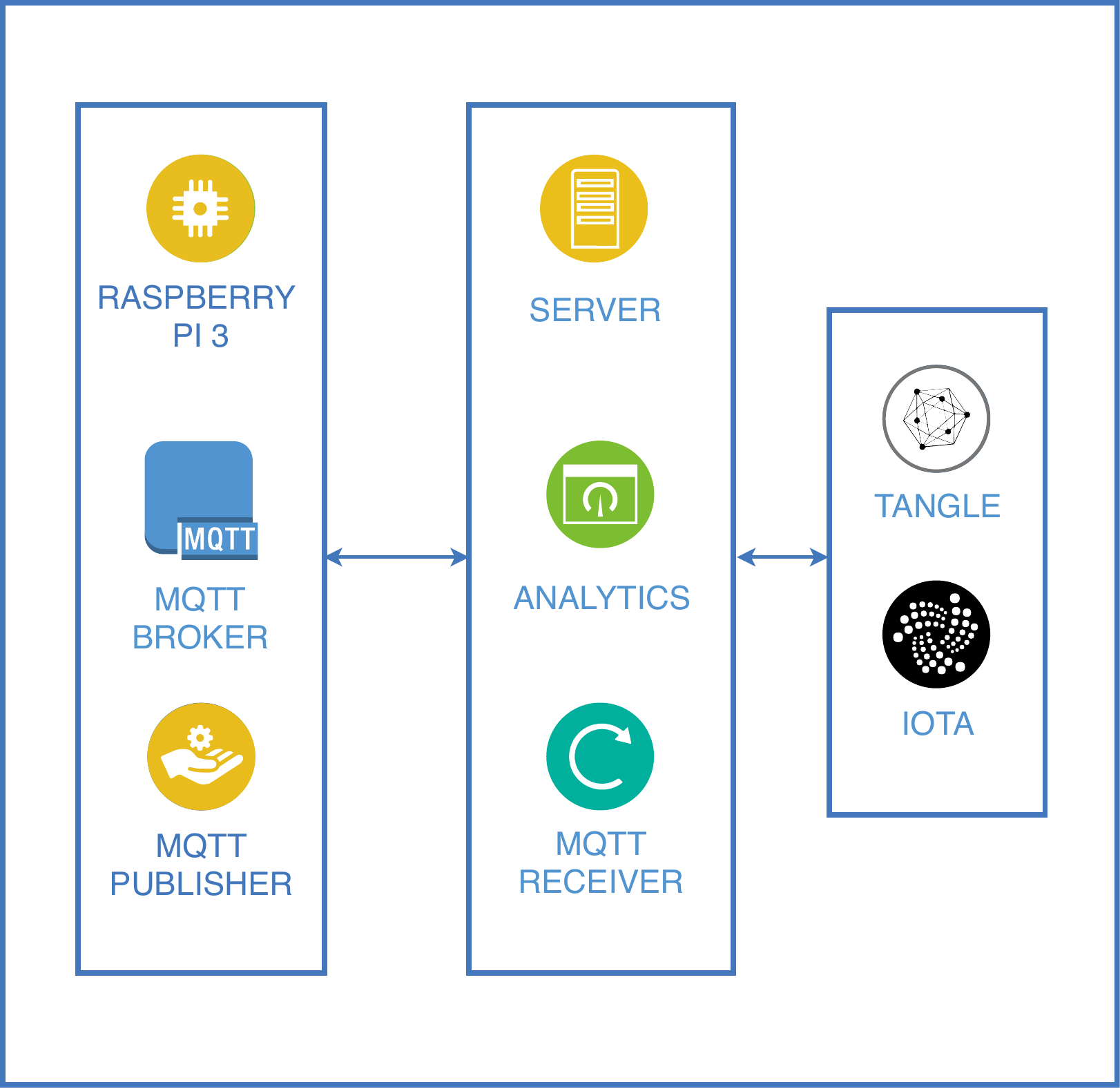}
\centering
\caption{High-level Overview}
\label{fig:highlevel}
\end{figure}

\subsection{Communication Architecture}

\begin{figure}
 \includegraphics[scale=0.3]{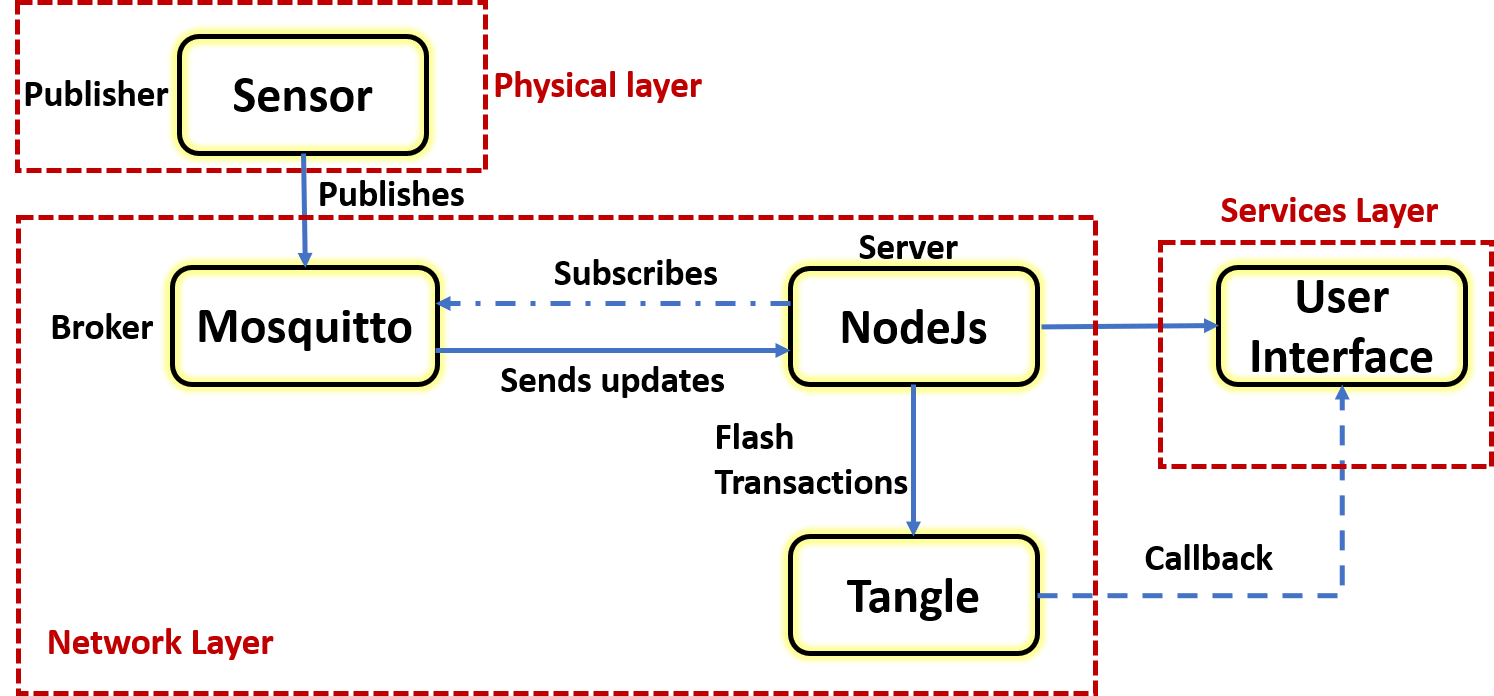}
 \centering
 \caption{Communication Architecture of POC Experiment}
 \label{fig:comarchpoc}
\end{figure}

Figure \ref{fig:comarchpoc} outlines the communication architecture of our POC. Physical layer contains the temperature sensor and the Raspberry Pi. To pass the temperature data recordings to higher level of our architecture, we used open-source and lightweight Mosquitto MQTT Broker\footnote{http://mosquitto.org/} which allows us to have the pure M2M MQTT communication. In addition, it supports Web Sockets as well as Secure Socket Layer (SSL) protocol, meaning that secure real-time updates are built-in.
PI also acts as a publisher, as it gathers the temperature data from the aforementioned sensor and publishes the results to the \textit{temperature:recording} topic. By subscribing to this topic, our server is able to get real-time updates. Once received by the Node.js server, temperature recordings are treated as services provided, and therefore, should be billed. We used IOTA's FC as a way to bill the client in real-time without any fees. Final balances of both parties are then saved to the Tangle, and upon the successful transaction, and UI is updated accordingly. 

\subsection{Results}
\begin{figure}
 \includegraphics[scale=0.10]{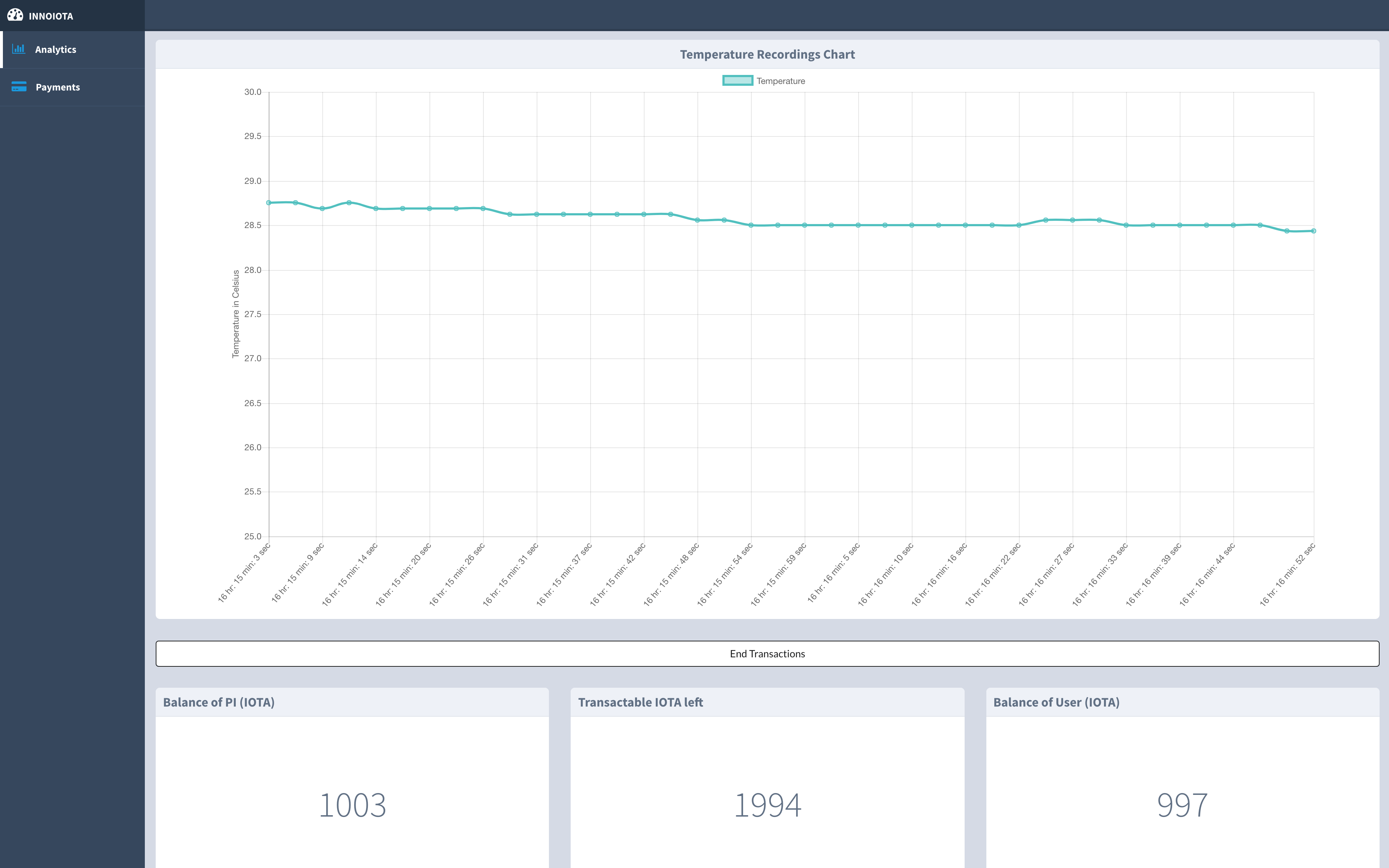}
 \centering
 \caption{User Interface for the Node Application}
 \label{fig:temperaturerecording}
\end{figure}

Figure \ref{fig:temperaturerecording} shows the working application in a web browser. Temperature recordings, as well as time stamps, are plotted in the chart in real-time, and the balances of both parties, and the amount of transaction-able IOTAs are shown underneath. This UI is communicating with the Node application through the Web Sockets and HTTPS calls for IOTA Flash transactions. 

\section{Conclusion}
\label{conclusions}
In this paper, we proposed a DLT-based charging and billing mechanism for EAVs. We use IOTA based payment system through machine-to-machine communication in order to carry out microtransactions for charging and billing in EAVs. To validate our framework, we developed a proof-of-concept with Raspberry Pi and a temperature sensor. The prototypical implementation is based on components providing the basis for a low-cost fast-deployable solution. Our proof of concept acts on a small scale with limited equipment that does not involve vehicles and charging stations. The plan is to prove the feasibility of the concept and attract the necessary support (including financial) to move to a full deployment. Our next steps include continuing the experimentation expanding the service layer, and then moving to a full-scale realization.






%



\bibliographystyle{IEEEtran}
\bibliography{bibliography}

\end{document}